\documentclass[prb,twocolumn,amsmath,amssymb,superscriptaddress,aps]{revtex4-1}
\usepackage{amssymb} 
\usepackage{amsmath}
\usepackage{mathtools}
\DeclarePairedDelimiter\abs{\lvert}{\rvert} 

\begin{document}

\title{First Principles Heisenberg Models of 2D magnetic materials: The Importance of Quantum Corrections to the Exchange Coupling} 

\author{Daniele Torelli}
\affiliation{Computational Atomic-scale Materials Design (CAMD), Department of Physics, Technical University of Denmark, DK-2800 Kgs. Lyngby, Denmark}
\author{Thomas Olsen}
\affiliation{Computational Atomic-scale Materials Design (CAMD), Department of Physics, Technical University of Denmark, DK-2800 Kgs. Lyngby, Denmark}

\date{\today}

\begin{abstract}
Magnetic materials are typically described in terms of the Heisenberg model, which provides an accurate account of thermodynamic properties when combined with first principles calculations. This approach is usually based on an energy mapping between density functional theory and a classical Heisenberg model. However, for two-dimensional systems the eigenenergies of the Heisenberg model may differ significantly from the classical approximation, which leads to modified expressions for exchange parameters. Here we demonstrate that density functional theory yields local magnetic moments that are in accordance with strongly correlated anti-ferromagnetic eigenstates of the Heisenberg Hamiltonian. This implies that density functional theory provides a description of these states that conforms with the quantum mechanical eigenstates of the model. We then provide expressions for exchange parameters based on a proper eigenstate mapping to the Heisenberg model and find that they are typically reduced by up to 17 \% compared to a classical analysis. Finally, we calculate the corrections to critical temperature for magnetic ordering for a previously predicted set of two-dimensional ferromagnetic insulators and find that the inclusion of quantum effects may reduce the predictions of critical temperatures by up to 7 \%. The effect is, however, predicted to be much higher for spin-1/2 systems, which are not included in the predictions of critical temperatures.
\end{abstract}

\maketitle

\section{Introduction}
The identification of ferromagnetism in a monolayer of CrI$_3$ in 2017\cite{Huang2017a} has initiated a vast interest in the field of two-dimensional (2D) magnetic materials.\cite{Gong2019, Gibertini2019, Sethulakshmi2019} From a technological point of view, 2D materials comprise a highly versatile platform for the design of electronics devices with properties tailored to specific applications. For example, bilayers of CrI$_3$ exhibit anti-ferromagnetic interlayer spin alignment, which may be switched to a ferromagnetic configuration by electrostatic gating. Thus bilayers of CrI$_3$ may act as efficient spin valves that can be controlled by a gate voltage and comprises and promising starting point for 2D spintronics applications.\cite{Song2019, Morell2018, Jiang2018a, Wang2018c} From a fundamental point of view the origin of magnetic order in 2D is distinctly different from its three-dimensional counterpart where magnetism is understood in terms of a spontaneously broken symmetry phase. In contrast, the Mermin-Wagner theorem\cite{MerminWagner} states that continuous symmetries cannot be broken {\it spontaneously} in 2D at finite temperatures and magnetic order thus crucially relies on magnetic anisotropy, which introduces an {\it explicitly} broken spin rotational symmetry. Since the discovery of magnetic order in monolayers of CrI$_3$, several materials have joined the family of 2D magnetic compounds. Most notably Fe$_3$GeTe$_2$,\cite{Fei2018} which is metallic (in contrast to CrI$_3$), FePS$_3$,\cite{Lee2016} which exhibits anti-ferromagnetic order, and bilayers of CrGeTe$_3$,\cite{Gong2017b} which becomes non-magnetic in the monolayer limit as a consequence of (weak) in-plane magnetic anisotropy. These materials are all characterized by a magnetic structure distinct from that of CrI$_3$ and have significantly expanded the range of possibilities for studying 2D magnetism. Considering the rapid evolution of synthesis techniques for 2D materials\cite{Zhou2018, Shivayogimath2019} it is expected that several new magnetic 2D materials will emerge in the near future. A wide range of new magnetic 2D materials have already been predicted from first principles calculations\cite{Mounet2018, Torelli2019c} and it remains to be seen whether any of these can be synthesized or exfoliated from bulk materials. However, such calculations only provide information about the magnetic ground state and additional modelling is required in order to predict whether or not magnetic order persist at finite temperatures\cite{Lado2017, Torelli2019}. In particular, a general framework for obtaining critical temperatures for magnetic order in 2D comprises a major challenge.

The theory of 2D magnetism is still in its infancy and standard approaches that work reasonably well in 3D is bound to fail in 2D. For magnetic insulators the thermodynamic properties is expected to be described accurately by Heisenberg models.\cite{Yosida1996} While such models are notoriously hard to solve, the thermodynamic properties at high temperatures are dominated by thermal fluctuations and quantum effects can be safely neglected such that a classical analysis suffices for obtaining critical temperatures\cite{Torelli2019} and exponents. This approach can be applied to real materials if the model parameters are obtained by an energy mapping between total energies obtained from first principles calculations and the energies obtained from the model in certain spin configurations.\cite{Xiang2013, Jacobsson2017, Kdderitzsch2002, Pajda2001, Olsen2017, Bose2010} Due to its simplicity the approach is always based on classical Heisenberg models, which works reasonably well for three-dimensional materials. However, in the case of 2D materials the anti-ferromagnetic configurations involved in the mapping are strongly correlated and are not necessarily well approximated by a classical configuration of the Heisenberg model.\cite{Yosida1996} The quantum corrections to the classical approach are inversely proportional to the magnitude of the spin and will be particularly prominent for systems with low spin.

In the present work, we show that density functional theory (DFT) applied to 2D magnetic insulators predict renormalized local magnetic moments that are in agreement with correlated eigenstates of the Heisenberg Hamiltonian and thus differ significantly from the classical prediction. This has two major implications: 1) It justifies the use of Heisenberg models for an accurate description of magnetic properties of insulators. 2) It shows that a proper energy mapping analysis should be based on the quantum states of the Heisenberg model. We will show that the latter point introduces a significant correction to the Heisenberg parameters compared to a classical analysis and this has crucial influence on the prediction of thermodynamic properties such as the critical temperature for magnetic order. We then apply the corrected exchange parameters to calculate critical temperatures for a previously predicted set of 2D ferromagnetic insulators. We also compare calculations  performed with and without Hubbard corrections and show that such corrections may modify the results significantly.

\section{Theory}
\subsection{Spin of the anti-ferromagnetic state}
We will consider the Heisenberg model with nearest neighbor interactions given by
\begin{align}\label{eq:H}
    H=-\frac{J}{2}\sum_{\langle ij\rangle}\mathbf{S}_i\cdot\mathbf{S}_j,
\end{align}
where the sum is over nearest neighbor (magnetic) atoms and $J$ is the exchange constant. For a bipartite lattice there is a unique (up to a SO(3) rotation in spin space) anti-ferromagnetic state where all sites are anti-aligned with neighboring sites. If $J<0$ this comprises the classical ground state (the Neel state) and for simplicity we will assume the spins to be aligned along the $z$-direction in this state. However, this is not an eigenstate of the quantum model and standard spinwave analysis shows that there is a state with lower energy referred to as the non-interacting magnon (NIM) state.\cite{Yosida1996} We will regard this as an approximate eigenstate of the Heisenberg Hamiltonian in the following. The NIM state has the same spin symmetry as the Neel state, but it is not an eigenstate of $S^z_i$. Instead the expectation value is given by
\begin{equation}\label{eq:spin}
    \langle S_z\rangle_{NIM}=S(1-\alpha/S),
\end{equation}{}
where the constant $S$ is the largest eigenvalue of $S_z$ and $\alpha$ is given by
\begin{align}
\alpha=\frac{1}{2}\bigg[\bigg\langle\frac{1}{\sqrt{1-|\gamma_\mathbf{q}|^2}}\bigg\rangle_{BZ}-1\bigg]
\end{align}
where
\begin{align}
\gamma_\mathbf{q}=\frac{1}{N_{nn}}\sum_{\Delta}e^{-i\mathbf{q}\cdot\mathbf{R}_\Delta}.
\end{align}
Here $\langle\ldots\rangle_{BZ}$ denotes a Brillouin zone average over $\mathbf{q}$ and $\mathbf{R}_\Delta$ are the $N_{nn}$ lattice vectors connecting nearest neighbor sites. The constant $\alpha$ is larger for low-dimensional models and thus becomes more important for 2D materials than for 3D materials. In the case of the honeycomb and square lattices the value of $\alpha$ can be evaluated numerically yielding 0.258 and 0.197 respectively.

Although the NIM state is obviously useful to describe properties of anti-ferromagnets ($J<0$), the derivation does not require that $J$ is negative. In particular, for a ferromagnetic Heisenberg model ($J>0$) the NIM state can be regarded as the approximate eigenstate of highest energy. With an accurate exchange-correlation functional, it is expected such a state should be represented by a configuration where the spins share the symmetry of the NIM state. This coincides with the spin symmetry of the Neel state and the spin configuration thus corresponds qualitatively to the classical anti-ferromagnetic configuration of interest. Since the associated spin densities are in principle accurately described by DFT, the ratio $m_{AFM}/m_{FM}$ should yield ($1-\alpha/S$) provided that the magnetic moments are localized. Ii is, however, far from obvious that standard approximations for the exchange-correlation functional will capture the intricate correlations in the anti-ferromagnetic state. In Sec. \ref{sec:results} we will provide evidence that a proper renormalization of the spin is captured in simple generalized gradient approximations exemplified by the Perdew-Burke-Ernzerhof (PBE)\cite{pbe} functional.

\subsection{Evaluating exchange constants}
For an $N$-site periodic bipartite lattice, the expectation value of the Hamiltonian \eqref{eq:H} using the Neel state is $NJS^2N_{nn}/2$ where $N_{nn}$ is the number of nearest neighbors. For an anti-ferromagnetic lattice ($J<0$) this comprises the classical ground state and for a ferromagnetic lattice ($J>0$) it is the classical state of highest energy. However, the NIM state has a lower(higher) energy for anti-ferromagnetic(ferromagnetic) models of the form \eqref{eq:H}. It is given by
\begin{align}
E^{NIM}=\frac{N}{2}(N_{nn}S^2J)\Big[1+\beta/S\Big],
\end{align}
where
\begin{align}
\beta=1-\bigg\langle\sqrt{1-|\gamma_\mathbf{q}|^2}\bigg\rangle_{BZ}.
\end{align}
For the honeycomb and square lattices the value of $\beta$ are given by 0.202 and 0.158 respectively. 

Exchange coupling constants are routinely evaluated from DFT using ferromagnetic and anti-ferromagnetic spin configurations in the simulations.\cite{Xiang2013, Jacobsson2017, Kdderitzsch2002, Pajda2001, Olsen2017, Bose2010} But it is usually assumed that such configurations can be mapped to the ferromagnetic state ($E^{FM}=-NJS^2N_{nn}/2$) as well as the \textit{Neel} state ($E^{Neel}=NJS^2N_{nn}/2$), which always leads to an overestimation of $J$. In exact DFT, the total energy of a given spin configuration should be mapped to the eigenstate of the Heisenberg model of the same spin symmetry. In particular, anti-ferromagnetic configurations have to be mapped to the NIM state, which provides a much better description of the anti-ferromagnetic state than the Neel state. For bipartite lattices this yields the expression
\begin{align}\label{eq:J_q}
J=\frac{\Delta E}{N_{nn}S^2(1+\beta/2S)},
\end{align}
where $\Delta E=E^{NIM}-E^{FM}$ is the energy difference per magnetic atom obtained from DFT. Again, we remark that the value of $\beta$ is in general smaller in 3D systems due to the 3D BZ average and it is often a better approximation to neglect the correction when evaluating exchange constants in 3D. However, as we will see below the inclusion of correlation effects in the energy mapping analysis can lead to significant corrections to the predictions of exchange constants and critical temperatures in 2D. Most spectacular are the effect on spin-1/2 systems on a honeycomb lattice where the exchange constants will be reduced be reduced by 17 \% compared to the classical prediction.

\subsection{Curie temperatures in 2D}
The model \eqref{eq:H} does not allow for magnetic order in 2D due to the Mermin-Wagner theorem\cite{MerminWagner} and one needs to consider models with terms that explicitly break the spin-rotational symmetry. Such terms originate from spin-orbit coupling and here we will assume that the most important effect on the magnetic order comes from single-ion anisotropy and nearest neighbor anisotropic exchange. The Hamiltonian then takes the form
\begin{align}\label{eq:H_ani}
    H=-\frac{J}{2}\sum_{\langle ij\rangle}\mathbf{S}_i\cdot\mathbf{S}_j-\frac{\lambda}{2}\sum_{\langle ij\rangle}S_i^zS_j^z-A\sum_i(S_i^z)^2,
\end{align}
where we have assumed isotropy in the $xy$-plane, which we take to comprise the atomic plane of the material. From hereon we restrict ourselves to the case of $J>0$. If one assumes that the easy-axis is along the $z$-direction, a simple spin-wave analysis then shows that the magnetic excitation spectrum has a gap given by
\begin{align}\label{eq:delta}
    \Delta=A(2S-1)+\lambda SN_{nn}.
\end{align}
A finite gap in the spectrum implies a broken spin-rotational symmetry and the model is expected to exhibit magnetic order at finite temperatures. If, however, the $z$-axis is not the easy axis one will obtain a negative spinwave gap ($\Delta<0$) and the spin-wave analysis is faulty because it has not been carried out on the magnetic ground state. If one assumes in-plane magnetic isotropy a negative spin-wave gaps thus implies the presence of a residual continuous rotational symmetry and the material cannot exhibit magnetic order by the Mermin-Wagner theorem. The sign of the spin-wave gap (evaluated from Eq. \eqref{eq:delta}) thus represents a descriptor for magnetic order at finite temperatures - even if it only corresponds to a physical quantity in the case of $\Delta>0$.

The anisotropy constants $\lambda$ and $A$ can be evaluated from DFT including spin-orbit coupling by considering energy differences between in-plane and out-of-plane spin configurations.\cite{Torelli2019c} However, the anisotropy terms in Eq. \eqref{eq:H_ani} will alter the expression for the exchange coupling and if spin-orbit coupling is strong the spin-orbit corrections to $J$ will typically be larger than the quantum corrections. For $S\neq1/2$ We thus propose to evaluate the Heisenberg parameters according to
\begin{align}
 A &=  \frac{\Delta E_{\mathrm{FM}}(1-\frac{N_{\mathrm{FM}}}{N_{\mathrm{AFM}}})+\Delta E_{\mathrm{AFM}}(1+\frac{N_{\mathrm{FM}}}{N_{\mathrm{AFM}}})}{(2S-1)S},\label{eq:A}\\
 \lambda &=  \frac{\Delta E_{\mathrm{FM}}-\Delta E_{\mathrm{AFM}}}{N_{\mathrm{AFM}}S^2},\label{eq:B}\\
 J &= \frac{ E_{A\mathrm{FM}}^{\parallel}- E_{\mathrm{FM}}^{\parallel}}{N_{\mathrm{AFM}}S^2(1+\beta/2S)},\label{eq:J}
\end{align}
where $\Delta E_{\mathrm{FM}(\mathrm{AFM})}=E_{\mathrm{FM}(\mathrm{AFM})}^{\parallel}-E_{\mathrm{FM}(\mathrm{AFM})}^{\perp}$ are the DFT energy differences between in-plane and out-of-plane magnetization for ferromagnetic(anti-ferromagnetic) spin configurations and $N_{\mathrm{FM}(\mathrm{AFM})}$ is the number of nearest neighbors with aligned(anti-aligned) spins in the anti-ferromagnetic configuration. Except for the quantum correction in Eq. \eqref{eq:J} and the factor of $(2S-1)$ in Eq. \eqref{eq:A} this comes out of an exact energy mapping to the classical Heisenberg model and the quantum corrections are simply added {\it ad hoc} to the expression for the exchange coupling. In addition we take $2S\rightarrow(2S-1)$ in the denominator of Eq. \eqref{eq:A} to ensure that the sign of the spinwave gap is strcitly determined by the sign of $\Delta E_\mathrm{FM}$. In the case of $S=1/2$ we take $A=0$. It should be stressed that in contrast to Eq. \eqref{eq:J_q} this is no longer rigorously derived since only two of the four spin configurations needed for the evaluation of Eqs. \eqref{eq:A}-\eqref{eq:J} are eigenstates of the quantum mechanical Heisenberg model. For example, with $J>0$ and out-of-plane easy-axis, the ground state of the Heisenberg model will be mapped $E_\mathrm{FM}^\perp$ and the eigenstate of highest energy will be mapped to $E_\mathrm{AFM}^\parallel$, but the remaining two spin configurations (which have energies obtainable from DFT) are not expected to be represented by eigenstates of the Heisenberg model.

The Curie temperature of the model \eqref{eq:H_ani} may be obtained by either classical Monte Carlo simulations\cite{Torelli2019} or a renormalized spin-wave analysis.\cite{Lado2017,Gong2017b,Torelli2019} We have previously shown that renormalized spin-wave theory breaks down in the case of large single-ion anisotropy\cite{Torelli2019} and in the present work we will evaluate Curie temperatures from classical Monte Carlo simulations. It may seem odd to rely on a classical analysis since we have argued that it is crucial to include quantum corrections when mapping DFT calculations to Heisenberg models. However, at temperatures in the vicinity of the critical temperature quantum fluctuations tend to be quenched by thermal fluctuations and a classical analysis becomes reliable even if they cannot be trusted at low temperatures. We note that spin-1/2 systems may comprise an important exception to this, since the single-ion anisotropy term becomes proportional to the identity in that case. It follows that magnetic order cannot exist at finite temperatures in spin-1/2 systems unless $\lambda\neq0$, which is in stark contrast to the predictions of classical Monte Carlo simulations where the value of $S$ simply introduces a rescaling of the Heisenberg parameters. For $S\neq1/2$, Monte Carlo simulations of the model \eqref{eq:H_ani} can be accurately fitted to a function of the form\cite{Torelli2019}
\begin{equation}\label{eq:tc}
T_\mathrm{C} = T_\mathrm{C}^\mathrm{Ising}f\left( \frac{\Delta}{J(2S-1)} \right )
\end{equation}
where 
\begin{equation}
f(x) = \tanh^{1/4} \left [  \frac{6}{N_{nn}} \log(1+\gamma x) \right]
\end{equation}
and $\gamma = 0.033$. $T_\mathrm{C}^{\mathrm{Ising}}$ is the critical temperature of the corresponding Ising model, which (in units of $JS^2/k_\mathrm{B}$) is given by 1.52, 2.27, and 3.64 for honeycomb, square and hexagonal lattices respectively.

\section{Results}\label{sec:results}
\subsection{Magnetic moments}
\begin{figure}[b]
	\includegraphics[width = 0.5\textwidth]{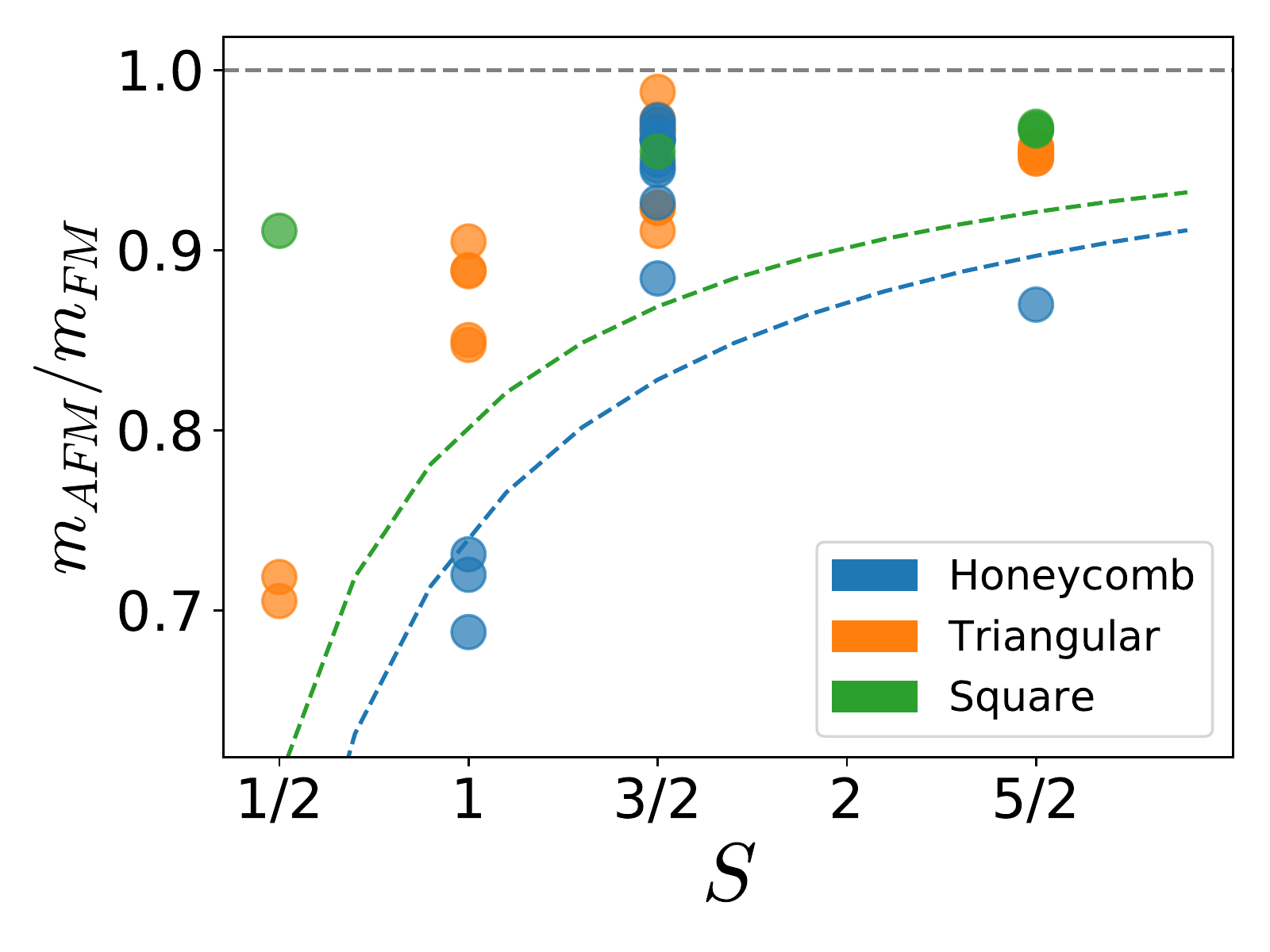}
	\caption{Ratio of average magnetic moments in ferromagnetic configurations for 51 insulating magnetic materials in C2DB. The dashed lines show the results obtained from a spin-wave analysis of the Heisenberg model with square and honeycomb lattices.}
	\label{fig:afm_corrections_magmom}
\end{figure}
In order to assess the performance of semi-local functionals for the correlated anti-ferromagnetic configuration of real materials, we have calculated the integral of the norm of the magnetization density
\begin{equation}
\label{eq:magmom}
m_{Abs} = \int \abs{m_{\uparrow}(\mathbf{r})-m_{\downarrow}(\mathbf{r})}d\mathbf{r}  
\end{equation}
for a wide variety of 2D materials in both the ferromagnetic and anti-ferromagnetic state. We have included all insulating magnetic materials present in the Computational 2D Materials Database\cite{Haastrup2018} (C2DB) where the magnetic atoms form either a honeycomb, square or triangular lattice. In this context, we define an insulator by a threshold of 0.2 eV for the band gap in both FM and AFM configurations to ensure that the basic electronic structure is not altered too much between the different spin configurations. The computational details can be found in Ref. \onlinecite{Haastrup2018}, but here we just mention that all calculations were carried out with the electronic structure code GPAW\cite{Enkovaara2010a, Larsen2017, Olsen2016a} using the PBE approximation\cite{pbe} for the exchange-correlation energy. Since the triangular lattice is not bipartite the classical state of lowest/highest energy is not collinear, but comprises a 120$^\circ$ non-collinear structure.\cite{Maksimov2019} However, as the far majority of DFT calculations in the literature (including energy mapping studies) are based on collinear structures we choose to base the energy mapping analysis of triangular lattices on a "striped" anti-ferromagnetic configuration where each site has two aligned and four anti-aligned nearest neighbors.\cite{Torelli2019c} Since this state does not comprise an extremum in the classical energy landscape it is not possible to perform a spinwave analysis to obtain the quantum corrections to the spin, but due to the four anti-aligned nearest neighbors the correction is expected to be similar to the square lattice.

In Fig. \ref{fig:afm_corrections_magmom} we display the ratio $m_{AFM}/m_{FM}$ (calculated from Eq. \eqref{eq:magmom}) per magnetic atom along with the theoretical predictions given by Eq. \eqref{eq:spin}. Although the calculations exhibit clear deviations from the Heisenberg prediction, there is a significant trend towards increasing quantum corrections with decreasing spin. It may be argued that a reduction of magnetic moments is expected in the anti-ferromagnetic state due to incomplete localization of the magnetic moments, which introduces cancellation of magnetization densities in the interstitial regions. However, such cancellation effects are expected to yield a ratio $m_{AFM}/m_{FM}$, which is {\it independent} of spin when averaged over a large class of materials. In contrast, Fig. \ref{fig:afm_corrections_magmom} shows a clear tendency towards decreasing staggered magnetization with decreasing spin, which is in accordance with the correlated state predicted by the Heisenberg model. 

\subsection{Exchange constants and critical temperatures}
The reduction of the ratio $m_{AFM}/m_{FM}$ with decreasing spin provides some confidence that DFT is able to capture (at least partly) the intricate correlations of the anti-ferromagnetic state at the level of generalized gradient calculations. This implies that exchange parameters calculated from energy mapping to the Heisenberg model should be corrected according to Eq. \eqref{eq:J}. This in turn may have a strong influence on the calculation of critical temperatures from the expression \eqref{eq:tc}. In Fig. 2 we show the quantum corrections to the exchange coupling parameters as well as the corrected critical temperatures relative to the classical estimates for all ferromagnetic insulators in the C2DB with $\Delta>0$. In the case of exchange parameters the reduction only depends on the value of $S$, whereas the corrected critical temperatures also depend on the spinwave gap $\Delta$. However, if $\Delta/J\ll1$ the change in critical temperature becomes $\delta T_\mathrm{C}/T_\mathrm{C}=3 \delta J/4J$ and the change in critical temperature thus largely follows the exchange parameter, which is also evident from Fig. 2. In general the reduction in critical temperatures is on the order of 5-7 {\%}. We note here that the non-corrected values of $J$ and $T_\mathrm{C}$ calculated here differ slightly from previously published results calculated with the same method\cite{Torelli2019c} since the effect of anisotropy is included in the evaluation of exchange constants, which were not the case in Ref. \onlinecite{Torelli2019c}. All the results are compiled in Tab. \ref{tab:data_nou}.
\begin{figure}[tb]
    \label{fig:corr_nou}
	\includegraphics[width = 0.5\textwidth]{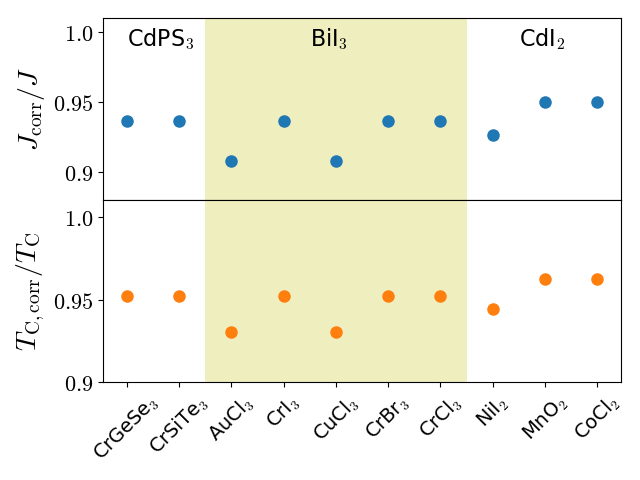}
	\caption{Corrections to exchange parameters (top panel) and Curie temperatures (lower panel) obtained with PBE for all insulating ferromagnets with positive values of $\Delta$ present in the C2DB. The three different regions signify the materials in the three prototypical structures CdPS$_3$, BiI$_3$ and CdI$_2$.}
\end{figure}
\begin{table}[tb]
  \begin{center}
\begin{tabular}{lccccc}
  Formula       & Prototype & $J_{\mathrm{corr}}$ [meV] & $\Delta$ [meV] & $S$ [$\hbar$] & $T_{\mathrm{C},corr}$ [K]           \\
    \hline
    \hline
NiI$_2$  &   CdI$_2$  &   7.84  &  0.97  &  1  &  84  \\
VSiSe$_3$  &   CdPS$_3$  &   29.1  &  0.079  &  1  &  59  \\
CuBr$_3$  &   BiI$_3$  &   9.11  &  3.5  &  1  &  64  \\
AuCl$_3$  &   BiI$_3$  &   8.32  &  1.44  &  1  &  48  \\
CuCl$_3$  &   BiI$_3$  &   13.6  &  0.048  &  1  &  30  \\
CoCl$_2$  &   CdI$_2$  &   1.99  &  0.068  &  3/2  &  29 \\
CrI$_3$  &   BiI$_3$  &   1.93  &  1.13  &  3/2  &  29  \\
CrBr$_3$  &   BiI$_3$  &   1.82  &  0.276  &  3/2  &  19  \\
MnO$_2$  &   CdI$_2$  &   0.514  &  0.434  &  3/2  &  17  \\
CrSiTe$_3$  &   CdPS$_3$  &   3.35  &  0.0136  &  3/2  &  14  \\
CrCl$_3$  &   BiI$_3$  &   1.31  &  0.041  &  3/2  &  9.3  \\
CrGeSe$_3$  &   CdPS$_3$  &   1.09  &  0.018  &  3/2  &  6.6
\end{tabular}
\end{center}
\caption{List of 2D magnetic insulating materials with positive exchange coupling $J_\mathrm{corr}$ (including quantum corrections) and positive spinwave gap $\Delta$ for PBE calculations. $T_\mathrm{C,corr}$ is the critical temperature evaluated from $J_\mathrm{corr}$ and $\Delta$ using Eq. \eqref{eq:tc}.}
\label{tab:data_nou}
\end{table}

Although we have argued that PBE (partly) captures quantum nature of anti-ferromagnetic configurations, a correct prediction of ground state energies for magnetic insulators pose a major challenge for DFT, since these materials typically involve strongly correlated electrons. This implies a large uncertainty in the prediction of exchange constants and critical temperatures. To quantify this we have repeated the calculations of the materials in the BiI$_3$ and CdI$_2$ prototypes using PBE+U where we have adopted the values of the Hubbard corrections (U) used in Ref. \onlinecite{Wang2006}. Tab. \ref{tab:data_plusu} shows a detailed list of materials obtained with the Hubbard correction and the calculated parameters. The overall trend in the corrections is very similar to Fig. 2 and are not shown. However the list of ferromagnetic candidates in Tab. \ref{tab:data_plusu} is rather different from Tab. \ref{tab:data_nou}. Several of the materials in Tab. \ref{tab:data_plusu} are lacking from Tab. \ref{tab:data_nou} and several materials present in Tab. \ref{tab:data_plusu} are absent from Tab. \ref{tab:data_nou}. Part of the reason is that we only include ferromagnetic insulators and the inclusion of the Hubbard correction may open a gap in a material that is predicted to be metallic without the Hubbard correction. In addition, the Hubbard correction may change the sign of both the spinwave gaps and exchange constants when U is included in the calculations. For example, with PBE+U the out-of-plane direction becomes a hard axis in Ni$_2$ and the material is predicted to have $T_\mathrm{C}=0$, but NiCl$_2$ acquires and out-of-plane easy with PBE+U and is thus predicted to have a finite critical temperature. In addition the Mn halides becomes ferromagnetic with PBE+U whereas they are anti-ferromagnetic with bare PBE. In Tab. \ref{tab:data_plusu} we have highlighted all the materials present in both cases with bold face. We note that the value of $T_\mathrm{C}=47$ K for CrI$_3$ obtained with PBE+U is in better agreement with the experimental value of 45 K compared to the bare PBE result of 29 K. This could, however, be coincidental since several other approximations enter the evaluation of $T_\mathrm{C}$ in the present framework. In addition, it has previously been shown that the value depends rather strongly on the choice of functional.\cite{olsen_mrs} In particular, the results of an LDA+U calculation gives a reduction of $J$ by 20 \% and there is no \textit{a priori} reason to believe that PBE should be better than LDA for the calculation of exchange constants. Nevertheless, the inclusion of Hubbard corrections typically provides a better description of the electronic structure in correlated materials, but it is not clear which value of U one should use.
\begin{table}[tb]
  \begin{center}
\begin{tabular}{lccccc}
  Formula       & Prototype & $J_{\mathrm{corr}}$ [meV] & $\Delta$ [meV] & $S$ [$\hbar$] & $T_{\mathrm{C}, corr}$ [K]           \\
    \hline
    \hline
\textbf{MnO$_2$}  &   CdI$_2$  &   6.1  &  0.186  &  3/2  &  87  \\
\textbf{CrI$_3$} &   BiI$_3$  &   3.6  &  1.29  &  3/2  &  47  \\
\textbf{CrBr$_3$}  &   BiI$_3$  &   2.62  &  0.197  &  3/2  &  23  \\
\textbf{CoCl$_2$}  &   CdI$_2$  &   0.34  &  0.36  &  3/2  &  12  \\
\textbf{CrCl$_3$}  &   BiI$_3$  &   2.05  &  0.0192  &  3/2  &  11  \\
FeI$_3$  &   BiI$_3$  &   0.22  &  0.31  &  5/2  &  9.3  \\
FeBr$_2$  &   CdI$_2$  &   0.125  &  0.25  &  2  &  8.1  \\
MnI$_2$  &   CdI$_2$  &   0.035  &  0.092  &  5/2  &  3.6  \\
NiCl$_2$  &   CdI$_2$  &   1.08  &  0.00041  &  1  &  2.7  \\
MnBr$_2$  &   CdI$_2$  &   0.028  &  0.031  &  5/2  &  2.3  \\
FeBr$_3$  &   BiI$_3$  &   0.031  &  0.126  &  5/2  &  1.7  \\
MnCl$_2$  &   CdI$_2$  &   0.020  & 0.0067  &  5/2  &  1.2
\end{tabular}
\end{center}
\caption{List of 2D magnetic insulating materials with positive exchange coupling $J_\mathrm{corr}$ (including quantum corrections) and positive spinwave gap $\Delta$ for PBE+U calculations. $T_\mathrm{C,corr}$ is the critical temperature evaluated from $J_\mathrm{corr}$ and $\Delta$ using Eq. \eqref{eq:tc}. The materials in bold face were also found with bare PBE and are present in Tab. \ref{tab:data_nou}.}
\label{tab:data_plusu}
\end{table}

\section{Discussion}
We have shown that DFT predicts a renormalization of localized magnetic moments in anti-ferromagnetic configurations of 2D insulators. The renormalization is in accordance with the predictions of the Heisenberg model and implies that energies of stationary states with anti-ferromagnetic spin alignment should be mapped to the corresponding correlated anti-ferromagnetic state of the Heisenberg model. This leads to a reduction in the predicted values of exchange parameters which in turn leads to a reduction of predicted Curie temperatures compared to an analysis based on classical Heisenberg models.

It would be interesting to compare the present approach to the method of infinitesimal rotations of local spin variables derived by Liechtenstein et al.\cite{LIECHTENSTEIN198765,LIECHTENSTEIN1985327} In that approach the magnetic force theorem is utilized to extract the exchange parameters from the ground state without any reference to different magnetic configurations. The methodology has the great advantage that all exchange parameters can be extracted without relying on magnetic configurations that may or may not comprise stationary states in DFT. Moreover, it can be argued that the inclusion of different magnetic configurations (as in the present work) introduces changes in the electronic structure (and associated changes in the energy) that may not be related to magnetic interactions whereas the Liechtenstein approach does not suffer from this problem. However, the method explicitly relies on the classical Heisenberg model and cannot include quantum effects at the level of magnetic interactions. In contrast, the stationary states of a given spin symmetry can naturally be regarded as eigenstates of the Heisenberg model and allows for a direct mapping to the quantum mechanical Heisenberg model. On the other hand, for metallic systems, there is typically a significant change in the electronic structure between different magnetic configurations and the Liechtenstein approach seems to be the only viable route to obtaining reliable exchange constants for metals.

For 2D ferromagnetic insulators the inclusion of quantum effects leads to corrections of $J$ and $T_\mathrm{C}$ the order of 5-10 \% for $S>1/2$. For $S=1/2$ the corrections may become significantly larger (17 \% reduction of the exchange coupling on the honeycomb lattice), but for these systems it is not straightforward to estimate the critical temperature for magnetic order and we have excluded them in the present study. We emphasize, however, that the errors originating from inaccuracies in DFT are likely to be somewhat larger than this. Nevertheless, the assumption of any first principles framework for evaluating magnetic interactions must be that the calculations are reliable and in that case any energy mapping approach must be based on the quantum mechanical Heisenberg model. Moreover, the corrections are easily expressed in an analytical form and can be included without any additional work. For more complicated lattices and exchange parameters beyond the nearest neighbor approximation the expression for the quantum corrections must be generalized, but this is straightforward to do for any given lattice. It should also be noted that this approach is not limited to 2D materials, but the corrections are in general larger compared to 3D materials.

The energy mapping scheme applied to obtain first principles Heisenberg models seems to provide an accurate and general framework for obtaining critical temperatures in ferromagnetic 2D insulators provided that DFT can deliver accurate energies of different spin configurations. It is, however, not obvious that DFT can do that with present day functionals and there is a strong need for a systematic assessment of functionals for such calculations. Moreover, the classical Monte Carlo approach is expected to fail for spin-1/2 systems and for such systems a full quantum mechanical treatment is required - either using numerical simulations (quantum Monte Carlo) or spin-wave theory beyond the random phase approximation. An even more important problem is the fact that no framework yet exist for evaluating critical temperatures in 2D metallic magnets. One possibility could be a generalization of the spin fluctuation theory developed by Moriya and Takahashi\cite{Moriya1978, Moriya1991} to 2D systems with magnetic anisotropy, but this is left to future work. 

\bibliography{references}

\end{document}